\documentclass[12Pt]{article}
\usepackage{amssymb,euscript,latexsym,amsmath}
\usepackage[dvips]{epsfig}
\usepackage{bm} 
\usepackage{enumerate} 
\usepackage{hyperref} 
\usepackage{color}
\usepackage{setspace}
\usepackage{subfigure}
\usepackage{bm}

\begin{document}

\title{Viscous evolution of point vortex equilibria:\\ The collinear state}
\author{Fangxu Jing, Eva Kanso and Paul K. Newton}


\maketitle

\begin{abstract}
We describe the viscous evolution of a collinear three vortex structure that corresponds initially to an inviscid point vortex fixed equilibrium, 
with the goal of elucidating some of the main transient dynamical features of the flow. Using a multi-Gaussian `core-growth' type of model, we show that  the system immediately begins to rotate unsteadily, a mechanism we attribute to a `viscously induced'
instability. 
We then examine in detail the qualitative and quantitative evolution of the system as it evolves toward the long-time asymptotic Lamb-Oseen
state, showing the sequence of topological bifurcations that occur both in a fixed reference frame, and in an appropriately chosen rotating
reference frame. The evolution of passive particles in this viscously evolving flow is shown and interpreted in relation to these evolving
streamline patterns.
\end{abstract}


\section{Introduction}

When point vortex equilibria of the 2D Euler equations (inviscid) are used as initial conditions for the corresponding Navier-Stokes equations (viscous), typically an interesting and complex dynamical process unfolds at short and intermediate time scales, which depends crucially on the details of the initial 
configuration. For long enough times,  Gallay \& Wayne proved recently that the Lamb-Oseen solution 
is an asymptotically stable attracting solution for all (integrable) initial vorticity 
fields~\cite{GaWa2005}. While very powerful, this asymptotic result does 
not elucidate the intermediate dynamics that take place in finite time and allow a given initial 
vorticity field to reach the single-peaked Gaussian distribution of the Lamb-Oseen solution. 
Given the rather large (and growing) literature on point vortex equilibria of the Euler equations, (see for example, ~\cite{Newton2001, ArNeStToVa2002, Aref2007}), we thought an analysis of how these equilibria evolve under the evolution of the full Navier-Stokes system would merit  a systematic treatment. Hence, 
in this paper, we begin an investigation of  the viscous evolution of a class of initial vorticity fields 
consisting of the superposition of $N$ Dirac-delta functions or point vortices~\cite{Newton2001}. Our initial configuration, shown in Figure \ref{fig:tripolesketch}, is  a collinear configuration of three point vortices, evenly 
spaced along a line (the $x$ axis), with strengths $2\Gamma$, $-\Gamma$, and  $2\Gamma$, respectively. Such a configuration, for the Euler equations, is known to be an {\em unstable} fixed equilibrium, as 
fleshed out most recently and comprehensively in~\cite{Aref2009}, but earlier in~\cite{TaTi1988}. We point out that this configuration, because of the strengths chosen
for each of the point vortices, is {\em not} what is commonly referred to as the `tripole' state~\cite{CaFlPo1989, HeKl1989, HeKlWi1991, KiKh2004} in which the vortex strengths sum to zero.
Our focus in this paper will be the  dynamics that takes place at the short and intermediate time scales, using this initial state in the Navier-Stokes equations,  
before the long-time asymptotic Lamb-Oseen solution dominates. This includes the dynamics of the surrounding passive field and the corresponding background time-dependent streamline pattern in an appropriately chosen reference frame which we argue is very helpful as a diagnostic tool to interpret the resulting flowfield.


\section{Problem Setting}

\begin{figure}[!t] 
	\begin{center}
\includegraphics[width=0.5\textwidth]{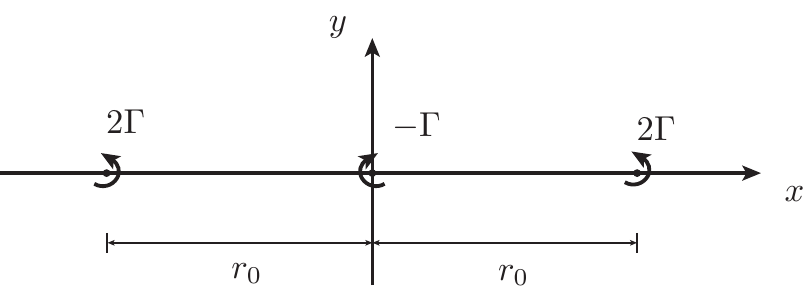} \caption{\footnotesize 
Fixed point-vortex equilibrium: three collinear and equally-spaced point vortices with the outer vortices of strength $2 \Gamma$ and the middle vortex of strength $-\Gamma$.} \label{fig:tripolesketch} 
	\end{center}
\end{figure}

Consider an incompressible fluid in an unbounded two-dimensional (2D) domain $\mathbb{R}^2$.
The fluid motion is governed by {\bf Navier-Stokes} equations,
written in terms of the vorticity field $\omega(\mathbf{x},t)$, a scalar-valued function 
of position $\mathbf{x}$ and time $t$, as follows
\begin{equation}
	\frac{ 
	\partial \omega}{ 
	\partial t} = -\mathbf{u}\cdot\nabla\omega + \nu\Delta\omega \, . \label{eq:ns} 
\end{equation}
The  kinematic viscosity $\nu$ is assumed to be constant.
The fluid velocity  $\mathbf{u}(\mathbf{x},t)$ is a vector-valued function of  $\mathbf{x}$ and $t$.
Both $\mathbf{x}$ and $\mathbf{u}$ are 
expressed in an inertial frame $\{\mathbf{e}_i\}_{i=1,2,3}$ where $(\mathbf{e}_1,\mathbf{e}_2)$ 
span the plane of motion, that is to say, one has
$\mathbf{x} = x\, \mathbf{e}_1 + y\, \mathbf{e}_2$ and  
$\mathbf{u} = u_x \, \mathbf{e}_1 + u_y \, \mathbf{e}_2$ or, equivalently, $\mathbf{x} = (x,y)$ 
and $\mathbf{u} = (u_x,u_y)$.  By definition, the vorticity vector 
$\boldsymbol{\omega} = \nabla \times \mathbf{u}$ is always 
perpendicular to the plane of motion and  can thus be expressed as
$\boldsymbol{\omega} = \omega \, \mathbf{e}_3$.
The velocity $\mathbf{u}$ and vorticity $\omega$ are related via the 2D 
Biot-Savart law
\begin{equation}
	\mathbf{u}(\mathbf{x},t) = \frac{1}{2\pi} \int_{\mathbb{R}^2} \frac{(\mathbf{x} - 
	\tilde{\mathbf{x}})^{\perp}}{\|\mathbf{x} - \tilde{\mathbf{x}}\|^2} \, \omega(\tilde{\mathbf{x}},t) 
	\, \text{d}\tilde{\mathbf{x}}\,, \label{eq:bs} 
\end{equation}
where  $\tilde{\mathbf{x}}$ is an integration variable and 
$\mathbf{x}^{\perp} = ( -y , x )$. Note that for 2D flows, the stretching  term 
$\boldsymbol{\omega}\cdot \nabla \mathbf{u}$ is identically zero thus 
does not appear in~\eqref{eq:ns} while the continuity equation ${\rm div}(\mathbf{u}) = 0$ 
is trivially satisfied when expressed in terms of vorticity.

The solution of the system of equations~\eqref{eq:ns} and~\eqref{eq:bs} depends, of course,  
on the choice of initial conditions $\omega(\mathbf{x}, 0)$. 
One solution of particular interest in this work is the well-known {\bf \em Lamb-Oseen solution} 
corresponding to a Dirac-delta initial condition $\omega(\mathbf{x}, 0) = \Gamma \delta(\mathbf{x})$, 
i.e., a point vortex placed at the origin  with  circulation or strength $\Gamma$ (more generally, 
the circulation $\Gamma$ around any closed curve $C$ in the fluid domain is defined as 
$\Gamma = \oint_C \mathbf{u}\cdot\text{d}\mathbf{s} = \int_A {\omega}\text{d}a$ and can 
be thought of as the flux of vorticity through the area $A$ enclosed by the curve $C$).  
Traditionally, the problem can be expressed compactly in complex notation with position 
variable $\mathbf{z}$,  
$\mathbf{z} = x + i y$ and $i=\sqrt{-1}$. The Lamb-Oseen solution is given by 
(see, for example,~\cite{Saffman1992, Newton2001} for more details)
\begin{equation}
	\omega(\mathbf{z},t) = \frac{\Gamma}{4\pi\nu t}\exp
	\left(-\frac{\|\mathbf{z}\|^2}{4\nu t}\right) \,. \label{eq:omegasingleoseen} 
\end{equation}
The corresponding velocity field $\mathbf{u}$ is given by, 
\begin{equation}
	\dot{\mathbf{z}}^* = u_x - i u_y = \frac{\Gamma}{2 \pi i} 
	\frac{1}{\mathbf{z}} \left[1 - \exp\left(-\frac{\|\mathbf{z}\|^2}{4 \nu t}\right)\right] \,, \label{eq:velsingleoseen} 
\end{equation}
where the dot notation $\dot{()} = {\rm d}()/{\rm d}t$ refers to the time derivative
and the notation $\mathbf{z}^\ast = x - i y$ refers to the complex conjugate.
According to \eqref{eq:omegasingleoseen}, the evolution of a vorticity field that is initially concentrated at the origin 
is such that the vorticity diffuses axisymmetrically as a Gaussian distribution. The spreading of the 
vorticity concentration can be quantified by the vortex {\bf \em{core}} or {\bf \em{support}} defined as $\rho = \sqrt{4 \nu t} = \sqrt{4\tau}$, where $\tau \equiv \nu t$. 

Despite the explicit, exact, and simple nature of the solution
(\ref{eq:omegasingleoseen}),  
 for more complicated initial data, explicit solutions of~\eqref{eq:ns} and~\eqref{eq:bs} are not analytically available 
for a general initial vorticity field. We are particularly interested in the viscous evolution of a class 
of initial vorticity fields $\omega(\mathbf{z},0) = \sum_{\alpha=1}^N \Gamma_\alpha\delta(\mathbf{z} 
- \mathbf{z}_\alpha)$ consisting of the superposition of $N$ Dirac-delta functions or point vortices.
Note that Gallay \& Wayne proved recently that the Lamb-Oseen solution 
is an asymptotically stable attracting solution for all (integrable) initial vorticity 
fields, see~\cite{GaWa2005}, of which $\omega(\mathbf{z},0) = \sum_{\alpha=1}^N \Gamma_\alpha\delta(\mathbf{z} 
- \mathbf{z}_\alpha)$ is a special case. 
It is  the dynamics that unfolds as the system evolves towards this final state that we are interested in.

The dynamics of $N$ point vortices in the plane is extensively analyzed in the context of the inviscid 
fluid model ($\nu = 0$), see for example~\cite{Newton2001} and references therein. 
The vorticity field remains then concentrated for all times at $N$ points whose position 
$\mathbf{z}_\alpha(t)$ ($\alpha = 1,\ldots,N$) is dictated by the local fluid velocity induced 
by the presence of the other vortices. The fluid velocity at an arbitrary point $\mathbf{z}$ in 
the plane that does not coincide with a point vortex is obtained from the Biot-Savart 
law~\eqref{eq:bs} which takes the form
\begin{equation}
	\dot{\mathbf{z}}^* = \sum_{\alpha = 1}^N \frac{1}{2 \pi i} 
	\frac{\Gamma_{\alpha}}{\mathbf{z} - \mathbf{z}_{\alpha}} \,, \label{eq:velnpv} 
\end{equation}
whereas the velocity at a point vortex $\mathbf{z}_\beta$ is given by subtracting the effect of that 
point vortex from~\eqref{eq:velnpv}, and replacing $\mathbf{z}$ with $\mathbf{z}_\beta$,  namely,
\begin{equation}
	\dot{\mathbf{z}}^*_{\beta} = \sum^{N}_{\alpha \neq \beta} \frac{1}{2 \pi i} 
	\frac{\Gamma_{\alpha}}{\mathbf{z}_{\beta} - \mathbf{z}_{\alpha}} \,. \label{eq:centervelnpv} 
\end{equation}
The $2N$ first-order ordinary differential equations~\eqref{eq:centervelnpv} dictating the inviscid
evolution of $N$ point vortices are known to exhibit regular, 
including fixed and moving equilibria, as well as chaotic dynamics depending on the number 
of vortices, their strengths and initial positions. The literature on this general topic is large, and we refer simply to the influential
1983 review article of Aref \cite{Aref1983}, along with the monographs of Saffman~\cite{Saffman1992} (especially Chapter 7)  and Newton \cite{Newton2001}
for an immediate entry into the literature. 
We also mention the 2008 IUTAM Symposium `150 Years of Vortex Dynamics' held at the Technical University of Denmark in which the lively state-of-the-art developments were reported \cite{Aref2010}.

In order to highlight the way in which the presence of viscosity affects the inviscid point vortex dynamics,  
we focus on studying the viscous evolution of $N$ point vortices whose location at time $t=0$ correspond 
to a fixed equilibrium of the inviscid point vortex model~\eqref{eq:centervelnpv}. 
For concreteness, we consider the case of $N=3$ collinear and equally-spaced point vortices as shown in 
Figure~\ref{fig:tripolesketch}.  Let  $r_0$ denote the distance between two adjacent vortices and let 
$2\Gamma$ be the circulation of the left and right vortices and  while~$-\Gamma$ 
be that of the center vortex. Also, let $\mathbf{z}_L, \mathbf{z}_C$ and $\mathbf{z}_R$ denote the positions of centers
of the left, center and right vortices, respectively. One can readily verify that this configuration
constitutes a fixed equilibrium of the inviscid point vortex model~\eqref{eq:centervelnpv}.

It is convenient to  non-dimensionalize the problem using the length scale $L=2r_0$  and 
the time scale $T$ dictated by $\Gamma$, namely, $T = L^2/\Gamma$. To this end, the 
non-dimensional parameters are given by 
\begin{equation}
	\tilde{r}_0 = \frac{1}{2}, \quad \tilde{\Gamma} = 1, 
	\quad \tilde{\nu} = \frac{\nu}{\Gamma}\,, 
\end{equation}
and the non-dimensional variables are of the form
\begin{equation}
	\tilde{\mathbf{z}} = \frac{\mathbf{z}}{L},\quad
	\tilde{\mathbf{u}} = \frac{\mathbf{u}}{L/T}, \quad \tilde{\omega} = \frac{\omega}{1/T}.
\end{equation}
For simplicity, we drop the $\tilde{}$ notation with the understanding that all parameters and variables are 
non-dimensional hereafter. The Navier-Stokes equation~\eqref{eq:ns} can be rewritten in dimensionless 
form
\begin{equation}
\frac{ 
	\partial \omega}{ 
	\partial t} = -\mathbf{u}\cdot\nabla\omega + \dfrac{1}{Re}\Delta\omega \, , \label{eq:ns2} 
\end{equation}
where $Re$ is the dimensionless Reynolds number, here defined as $Re = \Gamma/\nu$. 
Some words of caution are in order here, as we
will be comparing numerical simulations of the Navier-Stokes  equations with our model, and to do so 
requires that one is able to compare the direct numerical simulation (DNS) Reynolds number with the `model' Reynolds number. For this, it is better
to think of the Reynolds number as the ratio of inertial effects $-\mathbf{u}\cdot\nabla\omega$ over diffusive effects  $\Delta\omega$.
In some sense, one can think of the term $-\mathbf{u}\cdot\nabla\omega$ as being primarily responsible for the rotation we will discuss, while the term $\Delta\omega$ not only triggers the rotation, but diffuses the cores of the vortices. 
For  any  DNS, this creates an `effective' Reynolds number which depends not only on  the choice $\Gamma/\nu$, but also on numerical discretization since it affects the `rotation' and `diffusion'. 
The `model' Reynolds number
will be discussed in more detail in the upcoming sections.

\begin{figure}[!t]
\begin{center}
		\includegraphics[width=0.77\textwidth]{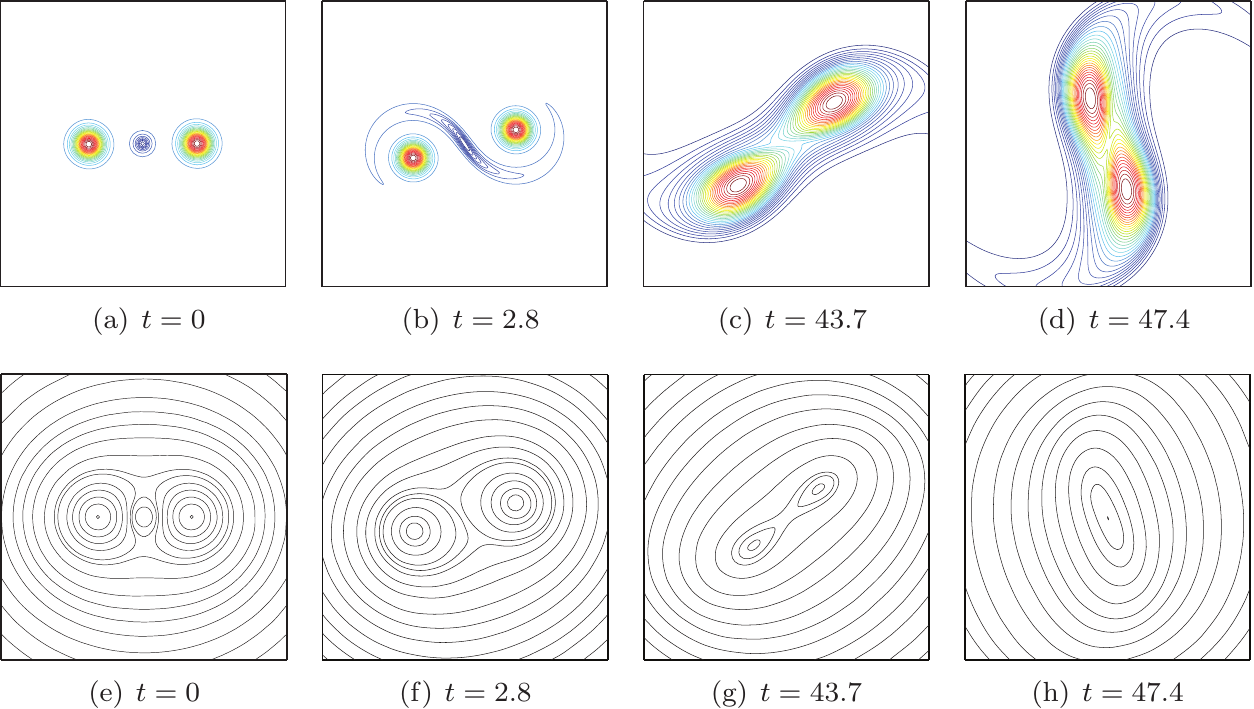}
		\caption {\footnotesize Vorticity contours (top row) and streamlines (bottom row) of Navier-Stokes simulation for $Re = 1000$ at $t=0, 2.8, 43.7$ and $47.4$. The vortex configuration rotates unsteadily for $t>0$. The center vortex stretches and diffuses out first, then the outer two vortices begin
to  merge. Eventually the vortex configuration approach a single Gaussian vortex.} \label{fig:vorticityandstreamlineNS}
		\end{center}
\end{figure}

By way of motivation, we first present a numerical solution of the system of equations~\eqref{eq:ns2} and~\eqref{eq:bs} subject to the initial condition $\omega(\mathbf{z},0) =  2\Gamma \delta(\mathbf{z} - r_0) - \Gamma \delta(\mathbf{z}) + 2\Gamma \delta(\mathbf{z} +  r_0)$. We use 
the numerical algorithm devised in~\cite{CoTa2008} that utilizes a second-order finite difference method with a multi-domain non-reflecting boundary condition to emulate the infinite fluid domain. This is a mesh-based method which poses a problem in handling the Dirac-delta initial conditions because they are not well-posed for discretization on a standard Euclidean mesh. 
To overcome this problem, we consider the initial conditions of the vorticity field as a superposition of three slightly diffused Gaussian peaks. 
In all the simulations presented here, we diffuse the initial Dirac-delta vorticity field by $\epsilon$ such that $\epsilon/Re = 2\times 10^{-4}$, e.g.,  $\epsilon = 0.2$ when $Re=1000$. 
This, of course, introduces a slight mismatch in the initial conditions used for the numerical simulation with those used in the model, an error we cannot completely eliminate, but should be kept in mind when comparing the simulation with the model.
We compute the time evolution of the vorticity field in the window $[-1.5, 1.5]\times[-1.5, 1.5]$
while the non-reflecting boundary conditions are imposed using the multi-domain technique with 10 nested domains, the largest of which is $2^{10}$ times the size of our result window.  The spatial and time steps are set to $\Delta x = \Delta y = 0.01$, $\Delta t = 0.02$.

Figure~\ref{fig:vorticityandstreamlineNS} depicts the time evolution of the vorticity contours (top row) and streamlines (bottom row) of Navier-Stokes solution for $Re = 1000$ at four time instants: $t=0, 2.8, 43.7$ and $47.4$. One notices that the vortex configuration begins to rotate unsteadily for $t>0$;  we refer to this motion as viscosity-induced rotation. One also notices that the center vortex stretches and diffuses out first, then the outer two vortices begin to merge. Eventually the vortex configuration approaches a single Gaussian vortex.


\section{The multi-Gaussian model}
\label{sec:model}

\begin{figure}[!t]
\begin{center}
		\includegraphics[width=0.77\textwidth]{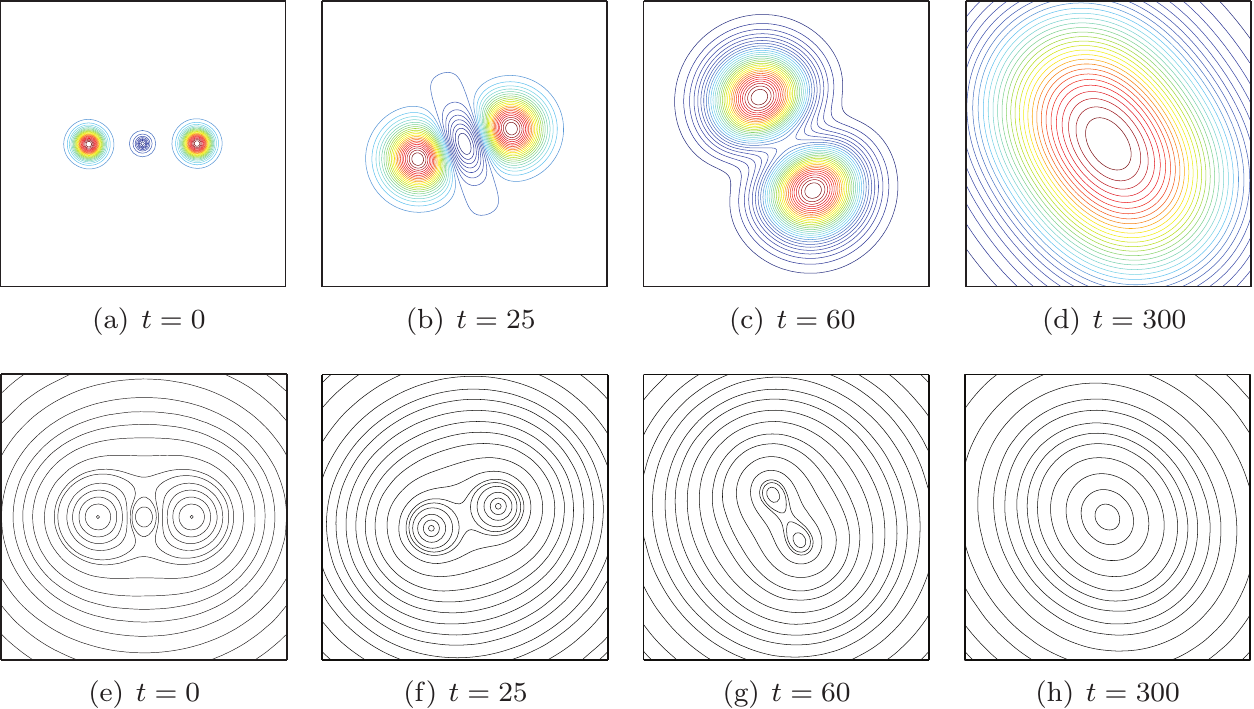} 
		\caption {\footnotesize Vorticity contours (top row) and streamlines (bottom row) of multi-Gaussian model for $\nu = 1/1000$ and $\Gamma = 1$ at four instants $t=0, 25, 60$ and $300$. Similar to Navier-Stokes simulation, the vortex configuration rotates unsteadily for $t>0$, the center vortex stretches and diffuses out first, then the outer vortices merge, eventually the vortex configuration approaches a single Gaussian vortex.} \label{fig:vorticityandstreamlinemodel}
		\end{center}
\end{figure}

In this section, we use a simple, analytically-tractable model to describe the dynamic evolution
of $N$ point vortices for non-zero (but small) viscosity ($\nu\neq 0)$. 
The model assumes that the vorticity of 
each initial point vortex spreads axisymmetrically as an isolated Lamb-Oseen vortex, thus modeling
the diffusion term $\nu \Delta \omega$ in~\eqref{eq:ns},
while its center moves according to the local velocity induced by the presence of the other (diffusing)
vortices, thus accounting for the convection term $\mathbf{u}\cdot\nabla \omega$  in~\eqref{eq:ns}.
It is worth noting here the recent work of Gallay \cite{Gallay2010} who 
analyses the inviscid limit $\nu \rightarrow 0$  of the 2D Navier-Stokes evolution of Dirac-delta initial conditions
and proves, under certain assumptions, that 
the solution of the Navier-Stokes equation converges, as $\nu \to 0$, to a superposition of Lamb-Oseen vortices. 
In this work, we show that, for small yet finite $\nu$, the multi-Gaussian model is able to capture qualitatively, though not quantitatively, some of the main features of the Navier-Stokes solution. 
Generally speaking, this class of models has been most highly developed in the numerical literature (see for example
\cite{CoKo2000, Leonard1980} and subsequent analysis in \cite{MaBe2002}) and is referred to as a `core-growth' class of models. One can trace the `splitting'  idea of the advection  and the diffusion
terms of the 2D Navier-Stokes equations, on which the core growth model is based,  at least back to Chorin's influential
paper~\cite{Chorin1973}, also used by Milinazzo and Saffman~\cite{MiSa1977}. 
In these papers, the diffusion was handled by a random walk approach.
Core growth models based on time-dependent solutions of the heat-equation were developed and used mostly by the numerical/computational vortex dynamics community, and are discussed and developed explicitly 
in \cite{Greengard1985, LuRo1991, KiNa1997, Rossi1996}.
In the context of numerical simulations,  focused studies can be found in the  works of 
Barba and Leonard \cite{Barba2004, BaLe2007},
and used in specific models in \cite{MeNe1990, NeMe1990, MeNeRaRu1995}. 
We mention, of course, also the works \cite{GaWa2002a, GaWa2002b, GaWa2005} and the 2009 Ph.D. thesis of 
Uminsky~\cite{Uminski2009} and follow-up work \cite{NaSaUmWa2009} which develops an eigenfunction expansion 
method based on the form of the heat-kernel. Additionally, we mention the body of work generated by Dritschel and co-workers, of which 
\cite{MaLeDr1994, RoDr2000} would be two relevant examples, whose aim is to elucidate via Lagrangian type numerical simulations, the
host of complex processes associated with mixing and dynamics in viscously evolving two-dimensional flows.

The model assumes that the vorticity field at all times is a superposition of multiple
Lamb-Oseen vortices
\begin{equation}
	\omega(\mathbf{z},t) = \sum_{\alpha = 1}^N \frac{\Gamma_{\alpha}}{4 \pi\nu t} \ 
	\exp\left(\frac{-\|\mathbf{z}-\mathbf{z}_{\alpha}\|^2}{4 \nu t}\right) \,. 
	\label{eq:omeganoseen} 
\end{equation}
The associated velocity field is computed by substituting \eqref{eq:omeganoseen} into \eqref{eq:bs}
\begin{equation}
	\dot{\mathbf{z}}^* = \sum_{\alpha = 1}^N \frac{\Gamma_{\alpha}}{2 \pi i (\mathbf{z} - \mathbf{z}_{\alpha})} \left[1 - \exp\left(\frac{-\|\mathbf{z}-\mathbf{z}_{\alpha}\|^2}{4 \nu t}\right)\right] \,. 
	\label{eq:velnoseen} 
\end{equation}
The velocity at the center  $\mathbf{z}_\beta$ of the $\beta^{th}$ vortex is given by subtracting the effect of that vortex 
and replacing $\mathbf{z}$ by $\mathbf{z}_\beta$ in~\eqref{eq:velnoseen}
\begin{equation}
	\dot{\mathbf{z}}^*_{\beta} = \sum_{\alpha \neq \beta}^N \frac{\Gamma_{\alpha}}{2 \pi i (\mathbf{z}_{\beta} - \mathbf{z}_{\alpha})} \left[1 - \exp\left(\frac{-\|\mathbf{z}_{\beta}-\mathbf{z}_{\alpha}\|^2}{4 \nu t}\right)\right] \,. \label{eq:centervelnoseen} 
\end{equation}
The system of equations in~\eqref{eq:omeganoseen}, \eqref{eq:velnoseen}, and~\eqref{eq:centervelnoseen} is referred to as the {\bf {\em multi-Gaussian}} model. 
We emphasize that we include in this model  the equation \eqref{eq:velnoseen} for the evolution of passive tracers in the `field' which is transported under the dynamics generated by ~\eqref{eq:omeganoseen} and~\eqref{eq:centervelnoseen}. This will be discussed more thoroughly in Section~\ref{sec:vorticity} and is relevant for comparisons of panels (a)-(d) of Figure~\ref{fig:vorticityandstreamlineNS} with (a)-(d) of Figure~\ref{fig:vorticityandstreamlinemodel}.

\begin{figure}[!t] 
	\begin{center}
	 \includegraphics[width=0.7\textwidth]{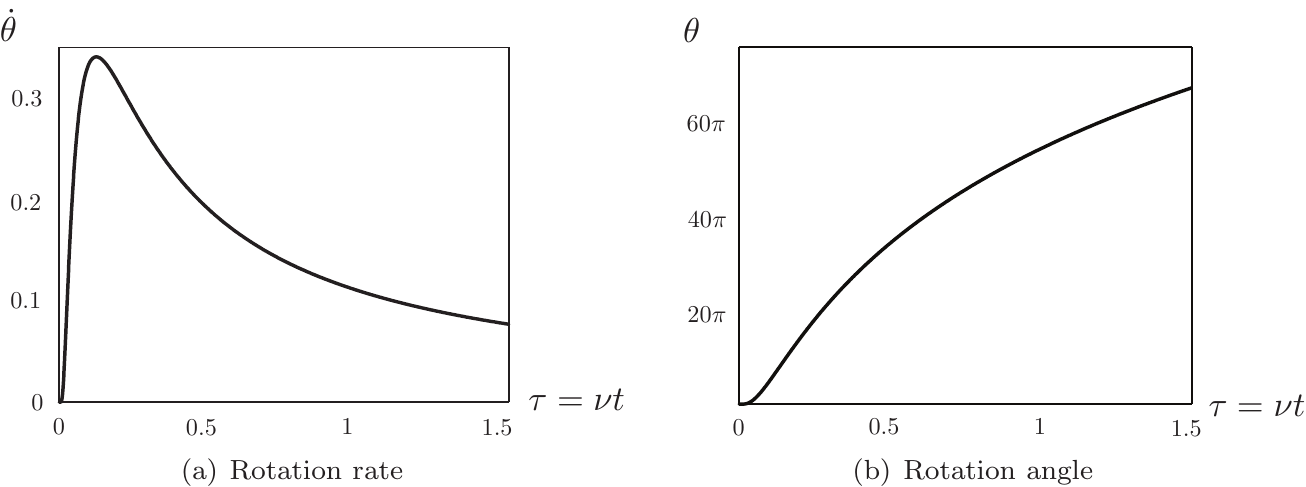}
	 \caption{\footnotesize Rotation rate $\dot{\theta}$ and rotation angle $\theta$ as functions of time $\tau = \nu t$ of multi-Gaussian model
for  $\Gamma = 1, r_0 = 0.5, \nu = 10^{-3}$.} 
\label{fig:thetadot2oseen}
	\end{center}
\end{figure}

According to~\eqref{eq:omeganoseen}, the vorticity field associated with the initial three-vortex configuration shown in Figure~\ref{fig:tripolesketch} is given by 
\begin{equation}
	\omega(\mathbf{z},t) = \frac{1}{4 \pi\nu t}\left[ 2\Gamma \exp\left(\frac{-\|\mathbf{z}-\mathbf{z}_L\|^2}{4 \nu t}\right) -\Gamma\exp\left(\frac{-\|\mathbf{z}-\mathbf{z}_C\|^2}{4 \nu t}\right) + 2\Gamma\exp\left(\frac{-\|\mathbf{z}-\mathbf{z}_R\|^2}{4 \nu t}\right)\right] \,. 
\end{equation}
The location of the centers of the vortices $\mathbf{z}_L$, $\mathbf{z}_C$ and $\mathbf{z}_R$ is obtained by solving the set 
of six first-order, ordinary differential equations in~\eqref{eq:centervelnoseen}. From symmetry, one can readily verify that 
$\dot{\mathbf{z}}_C^* = 0$ and that the centers of the vortices remain collinear and equally-spaced with constant distances for all time. 
The vorticity contours of~\eqref{eq:omeganoseen} are then plotted in Figure~\ref{fig:vorticityandstreamlinemodel} (top row).
The streamlines associated with the velocity field in~\eqref{eq:centervelnoseen} is shown in 
Figure~\ref{fig:vorticityandstreamlinemodel} (bottom row). Similarly to the Navier-Stokes solution depicted in 
Figure~\ref{fig:vorticityandstreamlineNS}, the dynamic evolution of the multi-Gaussian model is characterized by: 
(i) an unsteady rotation of the whole vortex  configuration for $t>0$, (ii) a stretching of the middle vortex, and 
(iii) eventual merging of the outer two vortices to form one single-peaked Gaussian of strength $3\Gamma$ 
as shown in Figure~\ref{fig:vorticityandstreamlinemodel}.
However, here some care is in order, as clearly Figures~\ref{fig:vorticityandstreamlineNS}(b)-(d) (DNS) and Figures~\ref{fig:vorticityandstreamlinemodel}(b)-(d) show some important differences.
Not only are the timescales different, but Figure~\ref{fig:vorticityandstreamlineNS}(b) shows a convective `wrappping' and `stretching'  of the middle vortex around the outer two before the diffusive effects kick in, whereas Figure~\ref{fig:vorticityandstreamlinemodel}(b) shows the stretching, but not the wrapping. Here it is important to remember that the
passively advected field, as shown in Figure~\ref{fig:passiveparticles}, is an important part of the model, and this field does show some of the same nonlinear wrapping features that appear in the DNS Figure~\ref{fig:vorticityandstreamlineNS}(b). One could say, in some respects, that the outer two vortices, being twice the strength of the inner one, are the primary `drivers' of the flowfield, which is perhaps why Figure~\ref{fig:vorticityandstreamlineNS}(e)-(h) match relatively well with Figure~\ref{fig:vorticityandstreamlinemodel}(e)-(h).
The `passively advected' inner vortex shown in Figure~\ref{fig:vorticityandstreamlineNS}(b) is better reflected in the passive particle field shown in Figure~\ref{fig:passiveparticles} and discussed 
at length in Section~\ref{sec:vorticity}. In turn, because the passively advected field in our model is not affecting the vorticity evolution, whereas in the DNS it is, this helps explain why the timescales associated with the two are different. The model is {\em not} an exact solution of the Navier-Stokes equations, and this appears to be the main physical manifestation of this fact.

The unsteady rotation rate of the vortex structure is obtained analytically as follows. 
From~\eqref{eq:centervelnoseen}, the velocity of  one of the outer vortices, 
say the right vortex, takes the form
\begin{equation}
	\dot{\mathbf{z}}_R^* = \frac{2\Gamma}{2 \pi i (\mathbf{z}_R - \mathbf{z}_L)} 
	\left[1 - \exp\left(\frac{-r_0^2}{\nu t}\right)\right] + \frac{-\Gamma}{2 \pi i 
	(\mathbf{z}_R - \mathbf{z}_C)} \left[1 - \exp\left(\frac{-r_0^2}{4 \nu t}\right)\right] \,. 
	\label{eq:velVR}
\end{equation}
Now, by symmetry one has ${\mathbf{z}}_L = -r_0 e^{i \theta}$, $\mathbf{z}_C = 0$ 
and ${\mathbf{z}}_R = r_0 e^{i \theta}$, where $\theta$ is the angle between the line traced 
by the vortex centers and the $x$ axis, and $\dot{\mathbf{z}}_R = ir_0 \dot{\theta}e^{i\theta}$. 
One gets, upon substituting into~\eqref{eq:velVR} and simplifying, that the rotation rate 
$\dot{\theta}$ is given by 
\begin{equation}
	\dot{\theta} =\frac{\Gamma}{2 \pi r_0^2} \left[\exp\left(\frac{-r_0^2}{4 \nu t}\right) - 
	\exp\left(\frac{-r_0^2}{\nu t}\right)\right]\,. \label{eq:thetadotoseen} 
\end{equation}
In Figure~\ref{fig:thetadot2oseen} is a depiction of $\dot{\theta}$ versus $\tau = \nu t$ 
which shows  that the rotation rate starts from zero, reaches a maximum value 
$\dot{\theta}_{max}$ at an intermediate time $\tau_{max} = \nu t_{max}$ and eventually decays 
to zero as $\nu t\rightarrow\infty$. The values of $\dot{\theta}_{max}$ and 
$ \tau_{max}$ are given by
\begin{equation}
	\tau_{max} = \nu t_{max} = \frac{3 r_0^2}{8 \ln 2} \approx 0.1353 , \qquad \dot{\theta}_{max} 
	= \frac{\Gamma}{2\pi r_0^2}\left[\exp\left(-\frac{2\ln 2}{3}\right) - \exp\left(-\frac{8\ln 2}{3}\right)\right] \,. 
\end{equation}
The orientation angle $\theta$ can be readily obtained
by integrating~\eqref{eq:thetadotoseen} in time
\begin{equation}
\theta = \frac{\Gamma}{2\pi \nu r_{0}^{2}}\left[\exp\left(-\frac{r_{0}^{2}}{4\nu t}\right)\nu t 
- \exp\left(-\frac{r_{0}^{2}}{\nu t}\right)\nu t + \frac{r_{0}^{2}}{4} 
\text{Ei}\left(-\frac{r_{0}^{2}}{4\nu t}\right) - r_{0}^{2} \text{Ei}\left(-\frac{r_{0}^{2}}{\nu t}\right)\right]\,, \label{eq:theta}
\end{equation}
where the exponential integral is defined as $\text{Ei}(x) = -\int_{-\infty}^{x} \exp(t)/t \ \text{d}t$ 
in the sense of principle value, which can be evaluated numerically to machine accuracy.

\begin{figure}[!t]
	\begin{center}
		\includegraphics[width=0.77\textwidth]{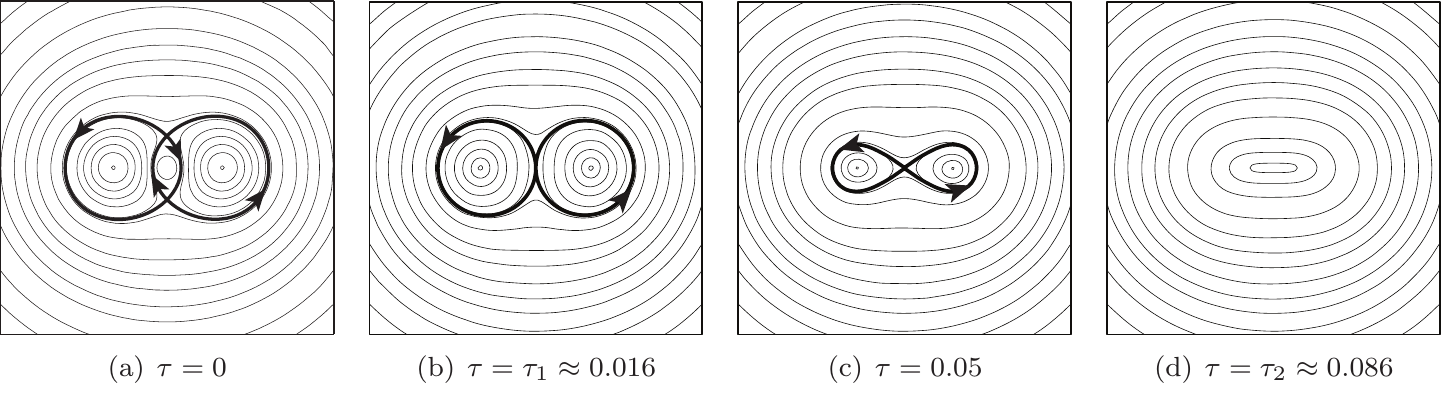}
				\caption {\footnotesize Evolution of the streamlines of the multi-Gaussian model. The separatrices are depicted in thick lines with arrows showing the direction of the flow. Instantaneous hyperbolic points are at intersections of separatrices while elliptic points are represented by circles.}\label{fig:separatrix} 
	\end{center}
\end{figure}
\begin{figure}
	[!t] 
	\begin{center}
	\includegraphics[width=0.6\textwidth]{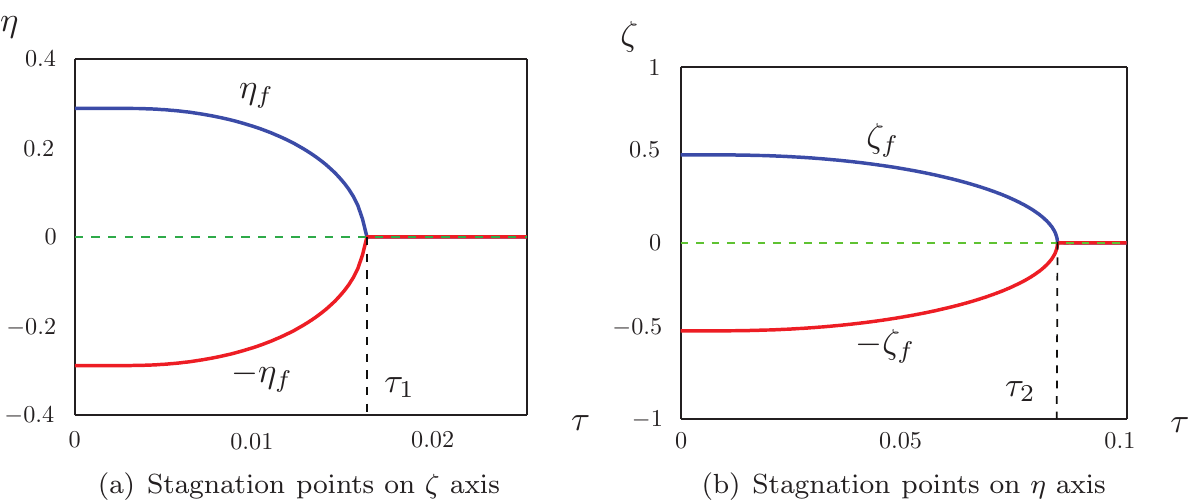}
		\caption{\footnotesize Instantaneous stagnation points: (a) a pair of hyperbolic stagnation point located at $(0,\pm\eta_{f})$
		for $\Gamma = 1, r_0 = 0.5$ and $\nu = 10^{-3}$. This pair collides at $(0,0)$
		at bifurcation time $\tau_1 \approx 0.016$. (b) a pair of elliptic stagnation points located at $(\pm\zeta_{f},0)$. This pair collides
		 with the now hyperbolic origin at bifurcation time $\tau_{2}\approx 0.086$.}
		 \label{fig:stagnationpoints} 
	\end{center}
\end{figure}

It is convenient for analyzing the flow to express the fluid velocity field 
$\dot{\mathbf{z}}$ in a frame  
co-rotating with the vortex configuration at the time-dependent 
rotation ${\theta}$. 
Let $\boldsymbol{\xi} = (\zeta,\eta)$ denote 
position of a point in the plane expressed in the rotating frame. 
The point transformation from  the rotating to the inertial frame is  given by
\begin{equation}
	{\mathbf{z}} = R \,{\boldsymbol{\xi}}\,, \quad 
	R = \left[ 
	\begin{array}{cc}
		\cos\theta & -\sin\theta\\
		\sin\theta & \cos\theta 
	\end{array}
	\right]\,, \label{eq:positiontransform} 
\end{equation}
The fluid velocity transforms as $\dot{\mathbf{z}} = R \mathbf{v}$ 
where $\mathbf{v} = (v_\zeta,v_\eta)$ is the velocity field expressed in the rotating frame.
In component form,  equations~\eqref{eq:velnoseen} transform as
\begin{equation}
	\begin{split}
		\frac{2\pi}{\Gamma}  v_{\zeta} \,= \,&  \frac{\eta}{\zeta^2 + \eta^2} \left[1 - \exp\left(-\frac{\zeta^2 + \eta^2}{4\tau}\right)\right]
		- \frac{2\eta}{(\zeta + r_0)^2 + \eta^2}\left[1 - \exp\left(-\frac{(\zeta + r_0)^2 + \eta^2}{4\tau}\right)\right]\\[2ex]
		& - \frac{2\eta}{(\zeta - r_0)^2 + \eta^2}\left[1 - \exp\left(-\frac{(\zeta - r_0)^2 + \eta^2}{4\tau}\right)\right]\,, 
	\end{split}
	\label{eq:velfield2a} 
\end{equation}
and
\begin{equation}
	\begin{split}		
		\frac{2\pi}{\Gamma} v_{\eta} \,= \,& -   \frac{\zeta}{\zeta^2 + \eta^2} \left[1 - \exp\left(-\frac{\zeta^2 + \eta^2}{4\tau}\right)\right]
		 + \frac{2(\zeta + r_0)}{(\zeta + r_0)^2 + \eta^2}\left[1 - \exp\left(-\frac{(\zeta + r_0)^2 + \eta^2}{4\tau}\right)\right]\\[2ex]
		& + \frac{2(\zeta - r_0)}{(\zeta - r_0)^2 + \eta^2}\left[1 - \exp\left(-\frac{(\zeta - r_0)^2 + \eta^2}{4\tau}\right)\right]\, ,
	\end{split}
	\label{eq:velfield2b} 
\end{equation}
where we used $\tau = \nu t$.

The instantaneous stagnation points of the velocity field (obtained by setting the right-hand side 
of (\ref{eq:velfield2a}), (\ref{eq:velfield2b}) to zero)
reveal important information about the  instantaneous streamlines of the fluid velocity field.
From symmetry of the velocity field, the instantaneous stagnation points must lie on the $\zeta$ and $\eta$ axes.  
One finds a total of five fixed points: one initially elliptic point at the origin, 
a pair of initially hyperbolic points at $(0,\pm\eta_{f})$ and a pair of initially elliptic points at $(\pm\zeta_f,0)$.  
The hyperbolic and elliptic character of these stagnation points is obtained by linearizing~(\ref{eq:velfield2a}) ,(\ref{eq:velfield2b})
about the instantaneous stagnation points and computing the eigenvalues of the linearized system. 
A pair of eigenvalues $\pm \lambda$ is associated with each stagnation point. 
One has a hyperbolic point if $\lambda$ is real and an elliptic point if $\lambda$ is pure imaginary.

At time $t=0$, the streamlines are those of the inviscid equilibrium, with a separatrix linking 
the two hyperbolic stagnations points on the $\eta$ axis, as shown in Figure~\ref{fig:separatrix}.
Initially, the separatrix divides  the fluid domain into four regions: three regions, 
one around each point vortex or elliptic point, and a fourth region 
bounded by the separatrix and the bound at infinity and void of point vortices. As time evolves, the location
of the instantaneous stagnation points change as shown in Figure~\ref{fig:separatrix}, 
and the separatrix evolves accordingly.
Note that the time-dependent separatrix does not constitute barriers to fluid motion and fluid particles typically 
move across this separatrix
as time evolves as discussed in more details in Section~\ref{sec:vorticity}.
Figure~\ref{fig:stagnationpoints} shows the coordinates of the stagnation points 
$\pm\eta_{f}$ and $\pm \zeta_{f}$ as functions of time. 
The pair of initially hyperbolic points $(0,\pm\eta_{f})$ start from $(0,\pm r_0/\sqrt{3})$ then collide together with the elliptic 
point at the origin in finite time $\tau_1 \approx 0.016$ to transform the origin into a hyperbolic point. This collision of 
instantaneous stagnation points is accompanied by a change in the streamline topology where the region around the 
center vortex disappears, see Figure~\ref{fig:separatrix}. Time $\tau_1$ is referred to as the first bifurcation time. (Note 
that the first bifurcation time $\tau_{1}$ does not correspond to when the cores of the three Gaussian vortices touch for the first time 
 which takes place at $\tau = r_{0}/16 = 0.015625$, nor does it correspond to when the cores of the two outer Gaussian vortices touch $\tau = r_{0}^{2}/4= 0.0625$. Indeed,  the definition of core size of a Gaussian function is rather 
ad hoc and bears little relevance to the dynamics of the multi-Gaussian model.)
Similarly,  $(\pm\zeta_f,0)$ starts from $(\pm r_{0},0)$ and collides at the now hyperbolic point at the origin at time 
$\tau_2 \approx 0.086$. Time $\tau_2$ is referred to as the second bifurcation time. For $\tau > \tau_2$, 
one has one single elliptic point at the origin as expected from the asymptotic Lamb-Oseen solution.

\section{Comparison to Navier-Stokes}

\begin{figure}[!t] 
	\begin{center}
	\includegraphics[width=0.9\textwidth]{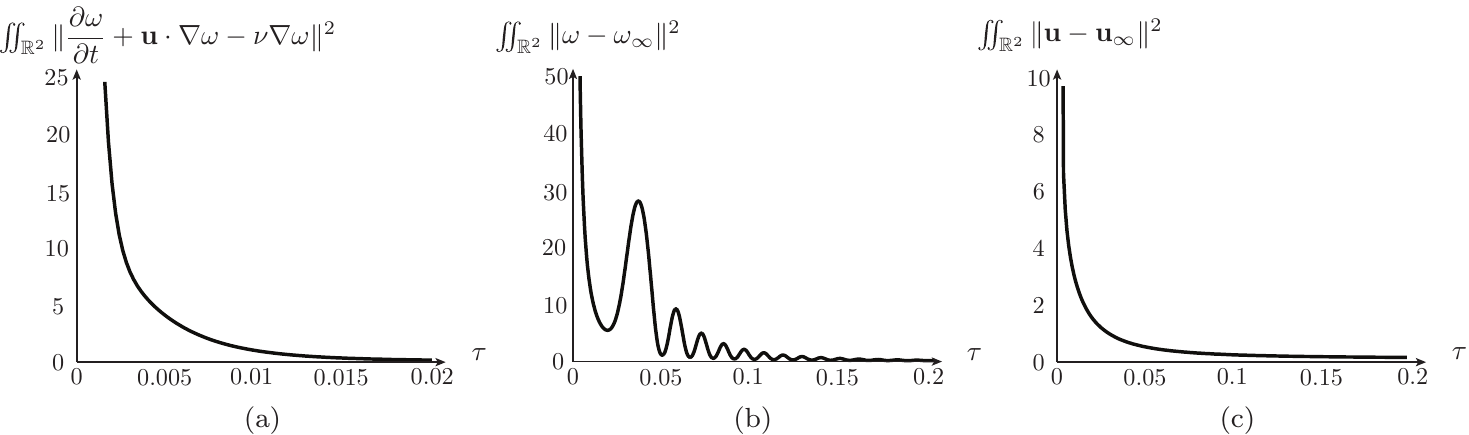}
\caption {\footnotesize  Multi-Guassian model:  (a) $L_2$ norm of residual $\sigma$ versus $\tau = \nu t$ for $\Gamma = 1$ and $\nu = 10^{-3}$; (b) $L_2$ norm of the difference in the vorticity field of the multi-Gaussian model and the single peaked Lamb-Oseen vortex with circulation $3\Gamma$; and (c) shows the difference in velocity field. Clearly, for long time, the model approaches the single peaked Gaussian but in short time, the multi-Gaussian, while not numerically accurate in comparison to the Navier-Stokes model as indicated in (a), its dynamics is richer than the single Gaussian as indicated in (b) and (c). 
} \label{fig:comparisonwithgaussian}
	\end{center}
\end{figure}

The residual $\sigma$ 
of the model is computed by substituting  the solution of 
\eqref{eq:omeganoseen} and \eqref{eq:velnoseen} into the Navier-Stokes equation~\eqref{eq:ns2}.
If the solution of the model is also an exact solution of the Navier-Stokes 
equation for a given set of initial conditions,  the residual $\sigma$ is identically $0$. 
In general, $\sigma$ is not zero (see discussions of this in \cite{Greengard1985}) and it can be viewed as an indication of the inaccuracy of the 
multi-Gaussian model. The $L_{2}$ norm of residual is plotted as a function of time 
$\tau=\nu t$ in Figure~\ref{fig:comparisonwithgaussian}(a) for the collinear vortex configuration considered here. 
Figure~\ref{fig:comparisonwithgaussian}(a) shows that as $\tau$ increases, the $L_{2}$ norm of $\sigma$
tends to zero, indicating that the multi-Gaussian model agrees with the Navier-Stokes solution
for $\tau$ large. From the result of Gallay \& Wayne~\cite{GaWa2005} and since the total circulation 
of the initial vorticity field is $3\Gamma \neq 0$, we know as $t\rightarrow \infty$ the 
Navier-Stokes solution approaches a single Gaussian vorticity distribution $\omega_{\infty} = ({3\Gamma}/{4\pi\nu t})\exp
\left(-{\|\mathbf{z}\|^2}/{4\nu t}\right)$ centered at the origin with circulation $3\Gamma$. 
We compute the difference between the multi-Gaussian
model and this asymptotic solution $\omega_{\infty}$. Figure~\ref{fig:comparisonwithgaussian}(b) 
and~(c) show the $L_{2}$ norm of the difference in both vorticity and velocity, respectively. These plots
confirm that the multi-Gaussian model approaches the asymptotic Lamb-Oseen solution for large time but at 
the intermediate times, the multi-Gaussian model exhibits richer dynamics  than the asymptotic 
Lamb-Oseen solution. While the dynamics of the multi-Gaussian model at these intermediate time
scales does not faithfully track the Navier-Stokes solution (as seen from Figure~\ref{fig:comparisonwithgaussian}(a)), it
does  capture more details than the asymptotic Lamb-Oseen vortex and its evolution
seems to exhibit the main qualitative features of the Navier-Stokes model as argued next.

\begin{figure}
	[!t] 
	\begin{center}
	\includegraphics[width=0.7\textwidth]{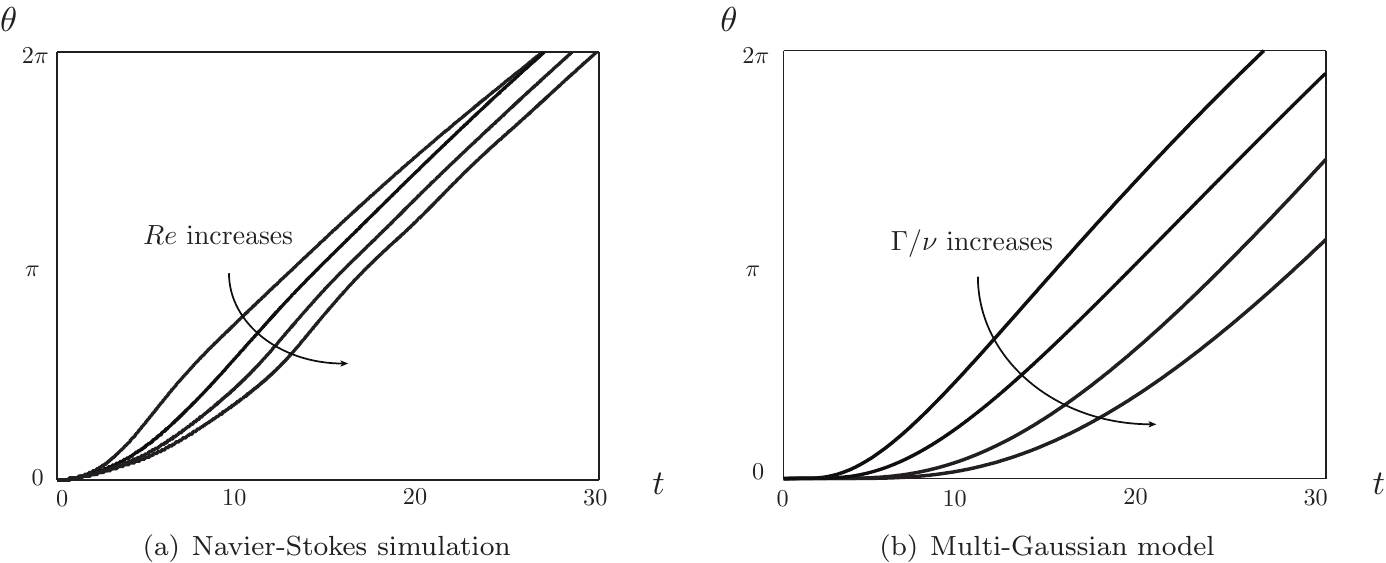}
		\caption {\footnotesize Comparison of rotation angle $\theta$ between (a)  the Navier-Stokes simulations and
		(b) the multi-Gaussian model. Navier-Stokes simulations are conducted with the same initial vorticity field for Reynolds numbers $Re=1000, 2000, 3000$ and $4000$. The results of the multi-Gaussian model are obtained for $\Gamma = 1$ and
		$\nu=1/100, 1/200, 1/300$, and $1/400$. The trend of both models is qualitatively similar.} \label{fig:theta_comparison} 
	\end{center}
\end{figure}

It is evident from Figures~\ref{fig:vorticityandstreamlineNS} and~\ref{fig:vorticityandstreamlinemodel} that both the Navier-Stokes equations 
and the multi-Gaussian model exhibit a viscosity-induced rotation as $t>0$. In Figure~\ref{fig:theta_comparison} we compare the qualitative 
trends of rotation angle $\theta$ obtained from the numerical solution to the Navier-Stokes equation and the analytical solution of multi-Gaussian model. In the Navier-Stokes solution, the rotation angle $\theta$ is obtained by computing the angle between the line traced by the vorticity peaks (see Figure~\ref{fig:vorticityandstreamlineNS}) and the $x$ axis while it is given by~\eqref{eq:theta} in the multi-Gaussian model. Clearly, both the Navier-Stokes solution and the model, although quantitatively distinct, exhibit similar qualitative trends in that the rotation angle $\theta$ is smaller when $Re$ increases (in Navier-Stokes) or equivalently when $\Gamma/\nu$ increases (in the model). As cautioned earlier about comparing DNS Reynolds numbers with the `model' Reynolds number, if both are 
thought of strictly as $\Gamma/\nu$, we can only claim qualitative overlap with the model and DNS. To obtain more quantitative overlap would require more effort on our part to obtain an accurate  Lagrangian based DNS to get a more detailed handle on the effective
`numerical' Reynolds number,  along with a modified model system that does more
to couple rotational effects with diffusive effects, neither of which are the immediate goals of the current work.

\begin{figure}
	[!t] 
	\begin{center}
	\includegraphics[width=0.45\textwidth]{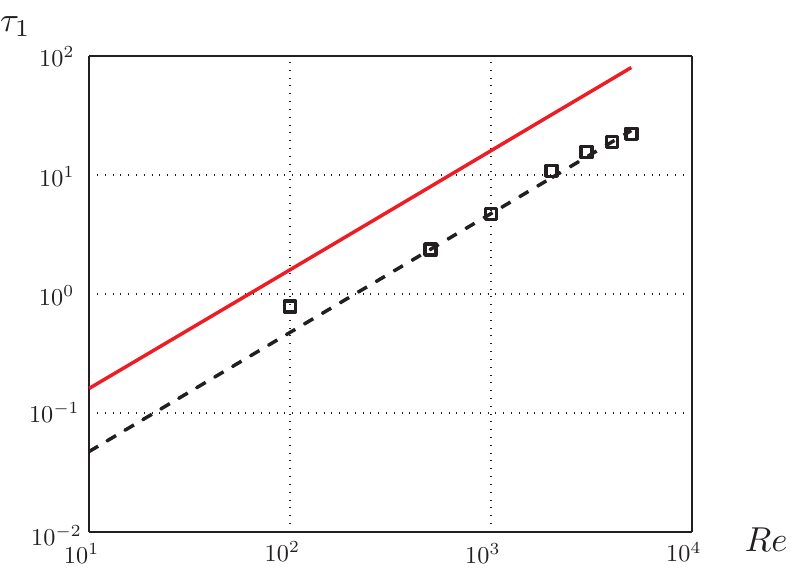} 
	\caption {\footnotesize Comparison of the times of the first bifurcation given by the Navier-Stokes simulations and the multi-Gaussian model in Log-Log plot. Simulation results are plotted as squares for $Re=100, 500, 1000, 2000, 3000, 4000, 5000$ and the dashed line is a fitted linear line obtained by least square distance rule. The solid line is the result from Multi-Gaussian model.} \label{fig:t1_Re} 
	\end{center}
\end{figure}

To quantify the difference between the Navier-Stokes solution and the multi-Gaussian model, we focus
on comparing  the first bifurcation time $\tau_1$ in the Navier-Stokes simulation for different $Re$ to
the first bifurcation time in the model. 
The result is plotted in Log-Log scale in Figure~\ref{fig:t1_Re} for $Re = 1000, 2000, 3000, 4000$ 
and $5000$ (plotted in squares). The dashed line is the best fitted straight line using the least square 
distance rule. The fitted line can be expressed as $\ln(\tau_{1}) = \ln(Re) - 5.356$, which means in 
linear scale, the fitting is $\tau_{1} = 0.00472 Re$. The simulation results are compared to the first 
bifurcation time $\tau_{1} = 0.016\Gamma/\nu$ as predicted by the multi-Gaussian model. While 
the first bifurcation in the simulations happens earlier than in the model, both the Navier-Stokes 
simulations and model indicate that $\tau_{1}$ is linearly dependent on $Re$ (or $\Gamma/\nu$).

\section{Evolution of vorticity in the multi-Gaussian model}
\label{sec:vorticity}

We now use the multi-Gaussian model to analyze
the fluid velocity and vorticity fields at intermediate time scales
before the asymptotic state of a single Lamb-Oseen vortex dominates.
The goal of this analysis is to understand the intermediate
mechanisms that lead the initial point vortex configuration
to reach the asymptotic Lamb-Oseen vortex.

\begin{figure}[!t]
\begin{center}
	\includegraphics[width=0.8\textwidth]{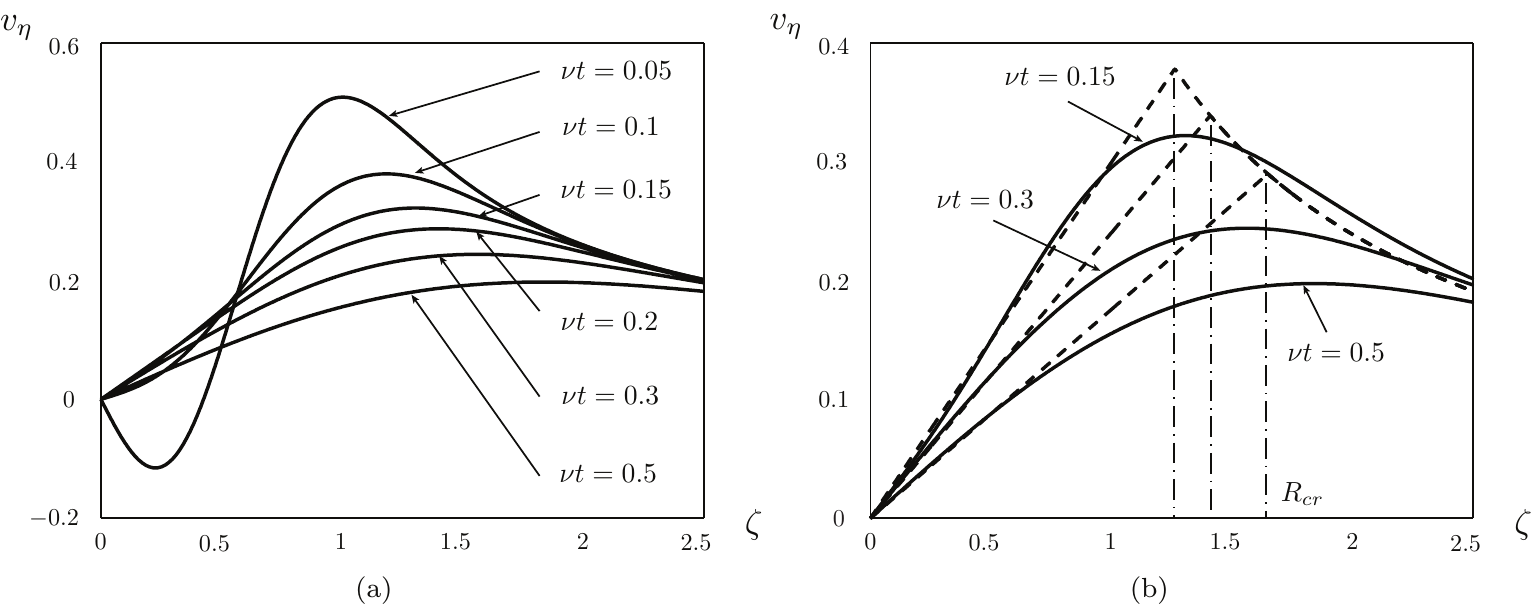}
\caption{\footnotesize The  velocity field induced by the collinear vortex structure becomes analogous
to that of a Rankine vortex. In particular, the component of velocity $v_{\eta}$ along the $\zeta$-axis is depicted.
In (a), we show the velocity profiles for $\nu t = 0.05, 0.1, 0.15, 0.2, 0.3$ and $0.5$. The maximum velocity
decreases as $\nu t$ increases. When $\nu t$ is small, e.g. $\nu t = 0.05$, $v_{\eta}$ is negative close to the origin.
This is because the vorticity is still relatively concentrated at the vortex centers.
In (b), we superimpose on the plots of $v_\eta$ versus $\zeta$ (solid lines)
the velocity of a Rankine vortex (dashed lines) with vorticity $3\Gamma$ and time-dependent core.
Clearly, the velocity field is similar to that induced by a rigid rotation close
to the origin and it is similar to an inverse decay at large distance from the origin.}
\label{fig:rankine}
\end{center}
\end{figure}

As time evolves, the vorticity field, initially concentrated at $\mathbf{z}_{C} = 0$ and
$\mathbf{z}_{L,R} = \mp1/2$, begins to spread spatially inducing a velocity field
similar to that of a  {\em Rankine vortex} with time-dependent core.
By way of background, the reader is reminded that the fluid velocity at a point $(\zeta,\eta)$
associated with a Rankine vortex at the origin with vorticity $3\Gamma$ (corresponding to
the total circulation of the collinear vortex structure) is perpendicular to the
distance $r$ from the origin and its value is given by
\begin{equation}\label{eq:rankinedef}
	v(r) = \left\{\begin{array}{lll}
		\dfrac{3\Gamma}{2\pi R_{cr}^{2}} \, r \, , & & \text{for} \ r\leq R_{cr}\\[4ex]
		\dfrac{3\Gamma}{2\pi} \dfrac{1}{r} \, ,  & & \text{for} \ r>R_{cr}
		\end{array}\right. \, \qquad (\text{here} \ r^2 = \zeta^2 + \eta^2) \, .
\end{equation}
The value $R_{cr}$ is referred to as the {\em core} of the Rankine vortex. For $r\leq R_{cr}$, the fluid velocity corresponds
to a {\em rigid rotation} while for $r>R_{cr}$, the velocity field decays proportionally to the inverse of the distance $r$.
As time evolves, the velocity field induced by the viscously evolving collinear vortex structure
becomes analogous to that of a Rankine vortex with vorticity  $3\Gamma$ and time-dependent core, as seen in Figures~\ref{fig:rankine}.
This analogy is especially evident in Figures~\ref{fig:rankine}(b) where we superimpose the velocity field of the Rankine vortex on that
induced by the viscously evolving collinear vortex structure at three different instances. Close to the origin, the velocity field of the collinear vortex structure looks like a rigid rotation, and the rotation rate is given by $\dot{\theta}$ in~\eqref{eq:thetadotoseen}. Since the rotation rate $\dot{\theta}$ is unsteady, the core size $R_{cr}$ of the Rankine vortex, obtained by equating $3\Gamma/2\pi R_{cr}^{2} = \dot{\theta}$, is time dependent and it increases with time $t$ as shown in Figures~\ref{fig:rankine}(a). As the distance from the origin increases, the velocity field of the collinear vortex structure decays analogously to the inverse decay with vorticity $3\Gamma$.

\begin{figure}
	[!t]
	\begin{center}
	 \includegraphics[width=0.65\textwidth]{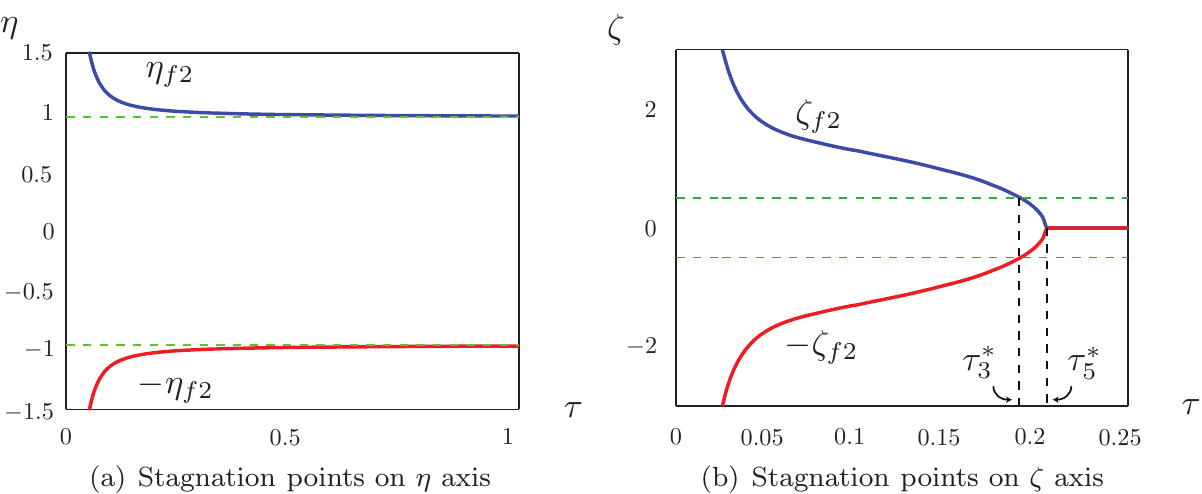}
	\caption{\footnotesize Relative velocity field: two new pairs of stagnation points appear from infinity as time $t>0$.
	(a) $\eta$-component of the pair of stagnation points $(0,\pm\eta_{f2})$.  This pair $(0,\pm\eta_{f2})$ eventually converge
	to $(0,\pm r_0 \sqrt{11/3})$ as $\tau \rightarrow \infty$, respectively.
	(b) $\zeta$-component of the pair of stagnation points  $(\pm\zeta_{f2},0)$. This pair $(\pm\zeta_{f2})$ reach $(\pm r_0,0)$ at
	bifurcation time $\tau_3^\ast \approx 0.1904$, and collapse at $(0,0)$ at bifurcation time $\tau_5^\ast \approx
	0.2045$. Parameters are $\Gamma = 1, r_0 = 0.5$ and $Re = 1000$.}	
	\label{fig:newstagnationpoints}
	\end{center}
\end{figure}

\begin{figure}
	[!t]
	\begin{center}
\includegraphics[width=\textwidth]{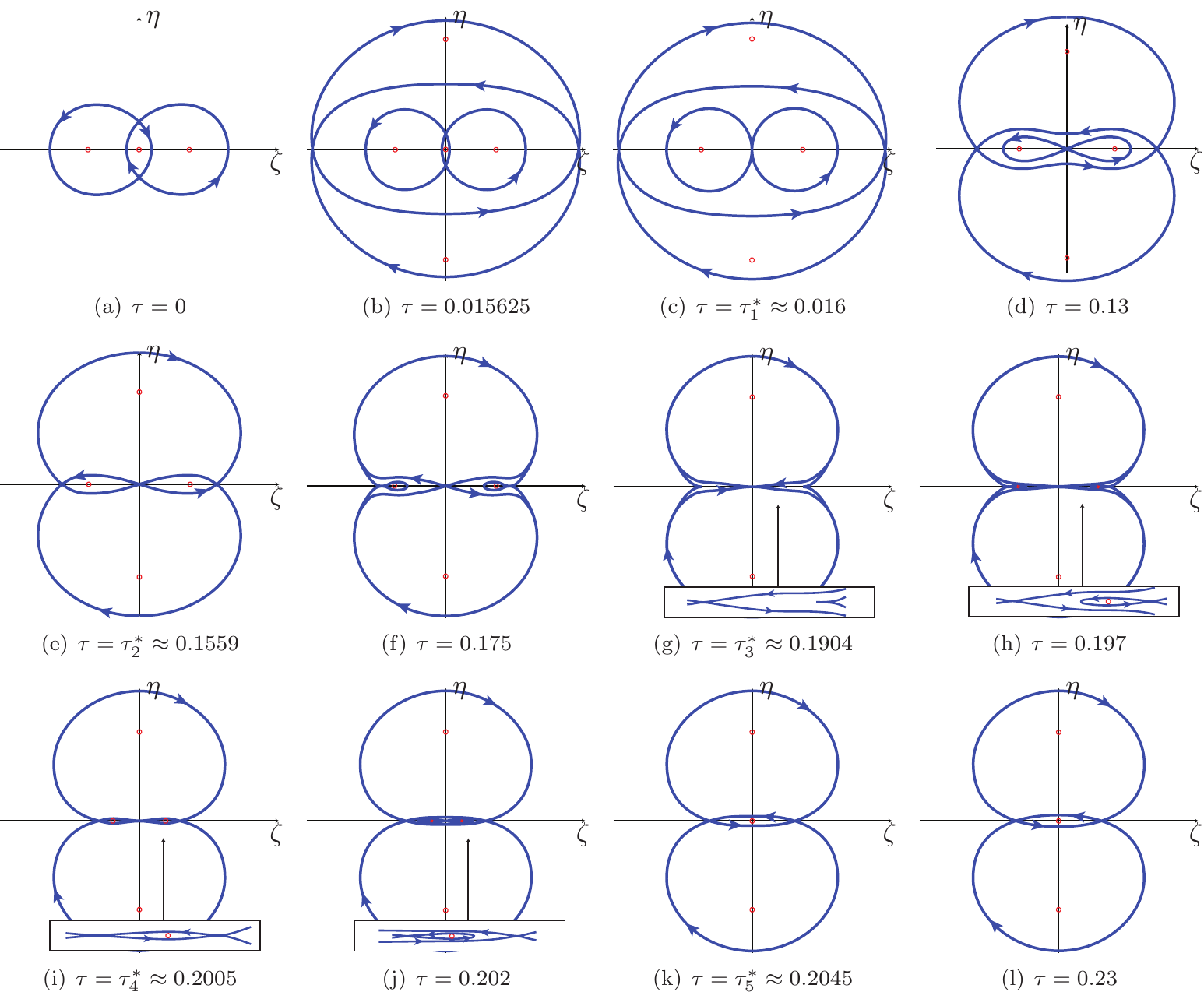} 
\caption {\footnotesize Evolution of the separatrices of the relative velocity field $\mathbf{v} - \dot{\theta} \boldsymbol{\xi}^\perp$. Instantaneous hyperbolic points are at intersections of separatrices and elliptic points are represented by circles. The outside separatrices and elliptic points in (b) and (c) are plotted out of scale. }
\label{fig:separatrix2}
	\end{center}
\end{figure}

Motivated by this analogy with the Rankine vortex, we examine the time evolution of the relative velocity field
\begin{equation}
\dot{\boldsymbol{\xi}} = \mathbf{v} - \dot{\theta} \boldsymbol{\xi}^\perp
\label{eq:relvel}
\end{equation}
obtained by subtracting a rigid body rotation from the fluid velocity field $\mathbf{v}$ expressed
in the rotating frame (written in~(\ref{eq:velfield2a}), (\ref{eq:velfield2b}) in component form).
Similarly to the analysis in Section~\ref{sec:model}, we identify the instantaneous stagnation points associated
with the relative velocity field~\eqref{eq:relvel}. Immediately as $t$ increases from zero ($t>0$),
in addition to the elliptic stagnation points located at the origin and $(\pm\zeta_f,0)$ and the hyperbolic points at $(0,\pm\eta_f)$,
one gets two new pairs of stagnation points appearing from infinity: one elliptic pair located at $(0,\pm\eta_{f2})$ and
one hyperbolic pair located at $(\pm\zeta_{f2},0)$.
Figure~\ref{fig:newstagnationpoints}(a) shows  the values of $\pm\eta_{f2}$ as functions of time. Clearly, $\pm\eta_{f2}$ start from $\pm\infty$ and
eventually converge to final values $\eta_{f2}|_{\tau\rightarrow\infty} = r_0\sqrt{11/3} \approx 1.9r_0$. The pair
$(0,\pm\eta_{f2})$ remains elliptic for all time.
Figure~\ref{fig:newstagnationpoints}(b) shows that the pair $(\pm\zeta_{f2},0)$ also starts from $\pm\infty$ and
reaches $\pm r_0$ at $\tau_3^\ast \approx 0.1904$ then the origin at $\tau_5^\ast \approx 0.2045$.
Meanwhile, the topology of the streamlines of the relative velocity field changes as a result of five
distinct bifurcations as depicted in Figure~\ref{fig:separatrix2} and explained next, thus the notation
$\tau_1^\ast, \ldots, \tau_5^\ast$.

The first bifurcation in the streamline topology is due to the same mechanism explained in Section~\ref{sec:model}
and takes place at the same time $\tau_1^\ast = \tau_{1}$, see Figure~\ref{fig:separatrix2}(c). The second bifurcation
does not coincide in time with the second bifurcation identified in Section~\ref{sec:model}, that is,
$\tau_2^\ast \neq \tau_2$. It is associated with a change in the streamline topology caused by a collapse of the
separatrices associated with the hyperbolic pair $(\pm\zeta_{f2},0)$ onto the separatrices of the now
hyperbolic point at the origin, see Figure~\ref{fig:separatrix2}(e). The third bifurcation occurs at $\tau_3^\ast$
when the hyperbolic points at $(\pm\zeta_{f2},0)$ collide with the elliptic points
at $(\pm \zeta_f = \pm r_0,0)$, respectively, causing them to change to hyperbolic points, see Figure~\ref{fig:separatrix2}(g).
After the third bifurcation, one still has two pairs $(\pm\zeta_{f2},0)$ and $(\pm \zeta_f,0)$ of stagnation points on
the $\zeta$-axis but with exchanged hyperbolic/elliptic characters.
The fourth bifurcation takes place at $\tau_4^\ast$ due yet to another collapse of the separatrices of the hyperbolic
point at the origin with the separatrices at the now hyperbolic points at $(\pm r_0,0)$, see
Figure~\ref{fig:separatrix2}(i). The fifth bifurcation takes place at $\tau_5^\ast \approx 0.2045$
when the now elliptic pair $(\pm \zeta_{f2},0)$ collides with the hyperbolic origin causing
it to turn into an elliptic point,  see Figure~\ref{fig:separatrix2}(k). This bifurcation sequence turns out to be
crucial in dictating the time evolution of the vorticity field which we visualize using colored passive
tracers as commonly done in  experimental and computational fluid mechanics
(see, for example,~\cite{HeKlWi1991}).

\begin{figure}
	[!t]
	\begin{center}
		\includegraphics[width=0.95\textwidth]{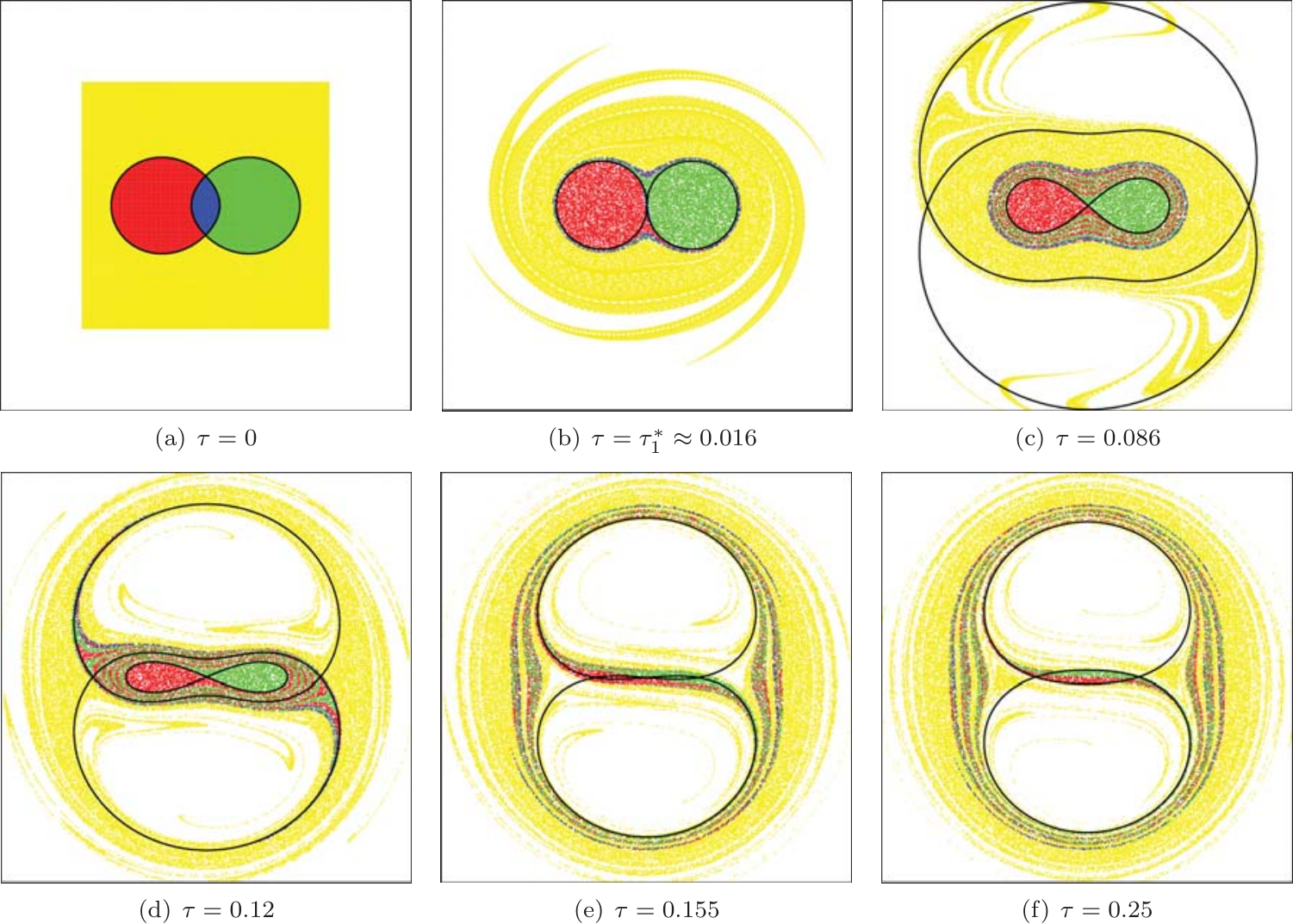}			
\caption {\footnotesize Colored passive tracers advected by the velocity field $\dot{\mathbf{z}}$ given in~\eqref{eq:velnoseen}
and depicted in the frame rotating with the vortex structure.
As time evolves, the passive tracers stretch and mix forming large lobes at a finite distance from the initial location of the vortex structure.
The separatices of the relative velocity field $\mathbf{v} - \dot{\theta} \boldsymbol{\xi}^\perp$ are superimposed in black at various instants in time.} \label{fig:passiveparticles}
	\end{center}
\end{figure}

We seed the flow at time $t=0$ with passive tracers of four different colors as shown in
Figure~\ref{fig:passiveparticles}(a) to distinguish the initial four fluid regions identified in
Section~\ref{sec:model}, namely, the three regions around the vortices bounded by the separatrix
(seeded with red, blue and green particles, respectively) and the fourth region (seeded with yellow particles)
bounded by the separatrix and the bound at infinity. We let the passive tracers be advected by the fluid velocity field
given in~\eqref{eq:velnoseen}. Snapshots of the passive tracers 
at six distinct instants in time are depicted in Figure~\ref{fig:passiveparticles}. 
As time evolves, the location of the stagnation points and
the associated separatrices change. Due to incompressibility, the particles initially in the region around
the middle vortex (blue color) `leak' along the unstable branch of separatrices associated with the instantaneous
hyperbolic points $(0,\pm\eta_{f2})$. At  $\tau_1^\ast$, Figure~\ref{fig:passiveparticles}(b) shows that all the particles are squeezed
out of the middle region. Meanwhile as time progresses, the fluid particles in yellow begin
to form lobes that stretch at a finite distance away from the initial location of the vortices, see Figure~\ref{fig:passiveparticles}(c).
Qualitatively, the passive tracers in Figure~\ref{fig:passiveparticles}(c) indicate a vorticity field similar to that obtained from the 
Navier-Stokes simulation in Figure~\ref{fig:vorticityandstreamlineNS}(d) (modulo the rigid rotation of the whole structure).
The formation of these lobes cannot be explained based on the analysis of the streamline patterns in Section~\ref{sec:model}.
Indeed, the formation of these lobes is initiated when the yellow passive tracers encounter the separatrices associated with
the hyperbolic points of the relative velocity field~\eqref{eq:relvel} $(\pm\zeta_{f2},0)$ that appear from infinity and move towards
the origin along the $\zeta$-axis (see Figure~\ref{fig:newstagnationpoints}(b)). The lobes then stretch and rotate
around the elliptic points $(0, \pm\eta_{f2})$ that appear from infinity and converge
to a finite distance away from the origin (see Figure~\ref{fig:newstagnationpoints}(a)).
Eventually, the passive particles initially placed in the regions around the point vortices, whose detailed
evolution is also dictated by the sequence of bifurfactions described in Figure~\ref{fig:separatrix2},
join the large lobes as well and begin to stretch and rotate at a finite distance away from the initial
vortex configuration, see Figure~\ref{fig:passiveparticles}(d)--(f). After the last bifurcation in Figure~\ref{fig:separatrix2}(k),
all the passive particles continue to rotate as shown in Figure~\ref{fig:passiveparticles}(f). We emphasize that this interesting dynamics of the passive particles,
which in turn indicates the evolution of the vorticity field, cannot be explained based solely on the analysis of the 
streamlines of the fluid velocity field of Section~\ref{sec:model}. In addition, because of the detailed and delicate
nature of the full series of topological bifurcations that  occur, to capture all but the first of these in a DNS would require considerable
further effort and is beyond the scope of the current manuscript.



\section{Conclusions}

The redistribution (inviscid) and diffusion (viscous) of delta-function initial distributions of vorticity, although configuration independent
for sufficiently {\em long} timescales, is highly dependent on the initial positions and strengths of the point vortices on short and intermediate timescales. These are typically the timescales in which much of the important mixing, transport, and redistribution of vorticity is achieved in many settings. Greengard's 1985 paper notwithstanding \cite{Greengard1985} pointing out that the types of models based on advection and core diffusion are not exact solutions of the Navier-Stokes equations, these ideas are exceptionally useful in getting a handle on some of the
important dynamical mechanisms that occur during the evolution towards the ultimate Lamb-Oseen state. 
In fact, one contribution of the current manuscript is to further quantify and understand the limitations of `core-growth' type models as 
diagnostic tools for understanding more and more complex flows and to point out some of the delicate issues in comparing
a DNS with these models.
Not surprisingly, core-growth type models are also useful as starting points for more sophisticated numerical methods which systematically exploit some of the main features 
\cite{CoKo2000, Rossi1996} (Also, see {\em Blobflow}, an open source vortex method package developed by Rossi, available at http://www.math.udel.edu/$\sim$rossi/BlobFlow as of October 2010). 

We summarize here with three main points associated with the viscous evolution of the three-vortex collinear state whose initial configuration
corresponds to an unstable inviscid fixed equilibrium:

(i) The presence of viscosity immediately `triggers' the underlying instability of the equilibrium, causing the vortices to rotate unsteadily;

(ii) In a fixed frame of reference, as the system evolves towards the ultimate Lamb-Oseen solution, the streamline patterns associated with the velocity field undergo a clear
sequence of topological bifurcations which we depict in Figure~\ref{fig:homotopic1}. We show the `homotopic equivalence' of each of the distinct patterns in the panels - the time and quantitative values of the pattern are not depicted, just the sequence of distinct patterns that
appear during the time sequence;

(iii) More interestingly, since the velocity field near the origin is of approximate solid-body (Rankine) form, if we subtract off this field and re-plot the homotopic sequence of patterns that emerges, shown in Figure~\ref{fig:homotopic2}, a far richer and more instructive sequence of 
patterns is revealed, one that is far more relevant for the understanding of the evolution of passive particle transport, as shown clearly  in Figure~\ref{fig:passiveparticles}.

We finish by mentioning connections of this work in two other contexts. First, there is by now a growing body of work on calculating
`time-dependent separatrices' in developing flows that goes under the name of `Lagrangian coherent structures' (LCS) \cite{Haller2001, HaYu2000}. Certainly these tools are potentially useful for further elucidating the intermediate timescale dynamics associated with the
evolution towards the Lamb-Oseen state, particularly for more complex initial patterns that perhaps start out as relative equilibria of the
Euler equations. 
Second, if one regards, the vorticity field as a probability density function associated, for example, with the positions of initial system of point vortices undergoing a random walk, there are meaningful interpretations of the models used in this paper that have been discussed most recently, for example, in \cite{AgVe1997, AgVe2001, Kevlahan2005}. While this interpretation
has not been the main focus of our work, we do find it potentially ripe for future development.


\paragraph{Acknowledgments.} This work is partially supported by the National Science Foundation through the CAREER award CMMI 06-44925 (FJ and EK) and the Grant NSF-DMS-0804629 (PKN).


\end{document}